\documentclass[11pt]{article}%
\usepackage{amsmath}
\usepackage{amsfonts}%
\usepackage{amssymb}
\begin{document}

\date{}
\title{\textbf{On Superfield Covariant Quantization in General Coordinates}}
\author{\textsc{D.M.~Gitman}$^{a)}$\thanks{E-mail: gitman@dfn.if.usp.br},
\textsc{P.Yu. Moshin}$^{a)}$\thanks{Tomsk State Pedagogical University, 634041
Tomsk, Russia; e-mail: moshin@dfn.if.usp.br}\thinspace\thinspace\ and
\textsc{J.L. Tomazelli}$^{b)}$\\ \\\textit{$^{a)}$Instituto de F{\'{\i}}sica, Universidade de S\~{a}o Paulo,}\\\textit{Caixa Postal 66318-CEP, 05315-970 S\~{a}o Paulo, S.P., Brazil}\\\textit{$^{b)}$Departamento de F\'{\i}sica e Qu\'{\i}mica, UNESP, Campus de
Guaratinguet\'{a}, Brazil}}
\maketitle

\begin{abstract}
We propose a natural extension of the BRST--antiBRST superfield covariant
scheme in general coordinates. Thus, the coordinate dependence of the basic
scalar and tensor fields of the formalism is extended from the base
supermanifold to the complete set of superfield variables. \vspace{0.5 cm}

\end{abstract}

\section{Introduction}

The principle of extended BRST\ symmetry applied to general gauge theories has
resulted in various schemes of covariant quantization \cite{BLT,3pl,mod3pl}.
It turns out that these schemes can be combined within the formalism
\cite{gln1,gln2,gl}, which realizes the modified triplectic algebra
\cite{mod3pl} in general coordinates. The differential operators $\Delta^{a}%
$,\emph{ }$V^{a}$,\emph{ }$U^{a}$ that form this algebra are constructed
\cite{gl} in terms of a nondegenerate antisymmetric tensor $\omega_{ij}$, a
symmetric tensor $g_{ij}$ and a scalar $\rho$, defined on a supermanifold with
a symmetric connection (Christoffel symbols). In Darboux coordinates, this
supermanifold (\emph{base supermanifold}) is parameterized by the fields and
antifields $\left(  \phi^{A},\bar{\phi}_{A}\right)  $ used in the quantization
schemes \cite{BLT,3pl,mod3pl}. It proves possible to fulfill the relations of
the modified triplectic algebra in case the tensor field $\omega_{ij}$ endows
the base supermanifold with a symplectic structure respected by the symmetric
connection (covariant derivative). In this respect, the base supermanifold can
be viewed as a Fedosov supermanifold \cite{gl}, which generalizes the notion
of Fedosov manifolds \cite{F}. The properties of such supermanifolds have been
recently studied in the papers \cite{gl2}.

In the original work \cite{gln1} on the modified triplectic quantization in
general coordinates,\ the authors raised the problem of a superfield
description of their formalism. This task calls for an extension of the
geometric contents of \cite{gln1,gln2,gl} to the complete set of variables of
\cite{BLT,3pl,mod3pl}, which can be regarded as superfield components in a
superspace with a pair of anticommuting coordinates \cite{L}. At present, two
different approaches \cite{gglm,gln3} to the mentioned problem have been
proposed. In \cite{gglm}, a superfield description of $\Delta^{a}$, $V^{a}$,
$U^{a}\ $is suggested, using a covariant differentiation in terms of
superfield variables. This formalism leaves intact the basic ingredients of
\cite{gln1,gln2,gl} as functions on the base supermanifold (accordingly
interpreted as a Fedosov supermanifold). On the contrary, in \cite{gln3} it is
proposed to extend the structure of a Fedosov supermanifold to the superfield
case. It turns out, however, that the resulting Christoffel symbols
\cite{gln3} cannot be regarded as connection coefficients having correct
properties under coordinate transformations. Thus, the approach \cite{gln3}
faces difficulties.

In this respect, the aim of the present work is to examine an alternative
extension of the superfield approach \cite{gglm}. Namely, we consider extended
counterparts of the fields $\omega_{ij}$, $g_{ij}$, $\rho$, defined on the
complete supermanifold of variables \cite{BLT,3pl,mod3pl}, and realize the
operators subject to the modified triplectic algebra in terms of such fields.
In the limit when the coordinate dependence of the mentioned fields is
restricted to the base supermanifold, one recovers the structure of a Fedosov
supermanifold, and the present quantization scheme reduces to that of
\cite{gglm}.

\section{Basic Objects}

In this section, we recall the basic ingredients of \cite{gglm}, namely, the
notion of a base supermanifold, the related construction of a \emph{triplectic
supermanifold}, and its superfield formulation.\textbf{ }In particular, we
remind the basics of tensor analysis on supermanifolds (for details, see the
monograph \cite{DeWitt} and papers \cite{gl,gl2}). We use DeWitt's condensed
notation \cite{condnot} and designations adopted in \cite{gglm}. Left-hand
derivatives are denoted by $\partial_{i}A=\partial A/\partial x^{i}$, and
right-hand derivatives are labeled by the subscript $``r"$, with the
corresponding notation $A_{,i}=\partial_{r}A/\partial x^{i}$. We assume that
covariant derivatives, $\nabla$, and other operators defined on tensor fields
\emph{act from the right:} $A\nabla$; besides, when necessary, the action of
an operator from the right is indicated by an arrow: $\overleftarrow{\nabla}$.
The Grassmann parity of a quantity $A$ is denoted by $\epsilon(A)$.

\subsection{Triplectic Supermanifold}

Let us consider a supermanifold $M$, $\mathrm{dim}\,M=N=2n$, with local
coordinates $(x^{i})$, $\epsilon({x}^{i})=\epsilon_{i}$. We now extend $M$ to
a supermanifold $\mathcal{M}$, $\mathrm{dim}\,\mathcal{M}=3N$, with local
coordinates $(x^{i},\theta_{a}^{i})$, where the additional coordinates
$\theta_{a}^{i}$ are combined into $Sp(2)$-doublets (labeled by the index
$a=1,2$) and possess the Grassmann parity, $\epsilon(\theta_{a}^{i}%
)=\epsilon_{i}+1$, opposite to that of $x^{i}$. We demand that the coordinates
$\theta_{a}^{i}$ transform as vectors under a change of coordinates on the
supermanifold $M$, indeed,\footnote{In \cite{gln1,gln2,gl}, the supermanifold
$\mathcal{M}$ is parameterized by coordinates $(x^{i},\theta_{ia})$, where
$\theta_{ia}$ transform as covectors, namely, ${\bar{\theta}}_{ia}=\theta
_{ja}\frac{\partial_{r}x^{j}}{\partial{\bar{x}}^{i}}$. Instead, in the paper
\cite{gglm} the parameterization $(x^{i},\theta_{a}^{i})$ is used, since it is
more convenient for a superfield formulation (see Subsection 2.2).{}}%
\[
\bar{x}^{i}=\bar{x}^{i}(x),\;\;\bar{\theta}_{a}^{i}=\theta_{a}^{j}%
\frac{\partial\bar{x}^{i}}{\partial{x}^{j}}\,.
\]

On the supermanifold $\mathcal{M}$, one defines a tensor field of type $(n,m)$
and rank $n+m$ as an object which in any local coordinate system $(x,\theta)$
is given by a set of functions $T_{\;\;\;\;\;\;\;\;\ j_{1}\ldots j_{m}}%
^{i_{1}\ldots i_{n}}(x,\theta)$ , with Grassmann parity $\epsilon
(T_{\;\;\;\;\;\;\;\;\ j_{1}...j_{m}}^{i_{1}\ldots i_{n}})=\epsilon
(T)+\epsilon_{i_{1}}+\cdot\cdot\cdot+\epsilon_{i_{n}}+\epsilon_{j_{1}}%
+\cdot\cdot\cdot+\epsilon_{j_{m}}$, that transform under a change of
coordinates $(x,\theta)\rightarrow(\bar{x},\bar{\theta})$ as a tensor field,
of the same rank and type, defined on the supermanifold $M$, namely,%
\begin{align*}
\bar{T}_{\;\;\;\;\;\;\;\;\ j_{1}...j_{m}}^{i_{1}...i_{n}}  &
=T_{\;\;\;\;\;\;\;\;\ k_{1}...k_{m}}^{l_{1}...l_{n}}\frac{\partial_{r}%
x^{k_{m}}}{\partial{\bar{x}}^{j_{m}}}\cdot\cdot\cdot\frac{\partial_{r}%
x^{k_{1}}}{\partial{\bar{x}}^{j_{1}}}\frac{\partial{\bar{x}}^{i_{n}}}{\partial
x^{l_{n}}}\cdot\cdot\cdot\frac{\partial{\bar{x}}^{i_{1}}}{\partial x^{l_{1}}%
}\\
&  \times(-1)^{\left(  \sum\limits_{s=1}^{m-1}\sum\limits_{p=s+1}^{m}%
\epsilon_{j_{p}}(\epsilon_{j_{s}}+\epsilon_{k_{s}})+\sum\limits_{s=1}^{n}%
\sum\limits_{p=1}^{m}\epsilon_{j_{p}}(\epsilon_{i_{s}}+\epsilon_{l_{s}}%
)+\sum\limits_{s=1}^{n-1}\sum\limits_{p=s+1}^{n}\epsilon_{i_{p}}%
(\epsilon_{i_{s}}+\epsilon_{l_{s}})\right)  }.
\end{align*}

Accordingly, a covariant derivative on $\mathcal{M}$ is defined as an
operation $\overset{\mathcal{M}}{\nabla}_{i}$ that maps a tensor field of type
$(n,m)$ into a tensor field of type $(n,m+1)$ and reduces to the usual
derivative $\partial_{r}/\partial x^{i}$ in a local Cartesian system on $M$.
Explicitly, the operation $\overset{\mathcal{M}}{\nabla}_{i}$ has the form of
a $\theta$-extension of the covariant derivative $\overset{M}{\nabla}_{i}$ on
the supermanifold $M$,%
\begin{equation}
\overset{\mathcal{M}}{\overleftarrow{\nabla}}_{i}=\overset{M}{\overleftarrow
{\nabla}}_{i}-\frac{\overleftarrow{\partial}}{\partial\theta_{a}^{k}}%
\theta_{a}^{m}\overset{M}{\Gamma}\,_{\;\;mi}^{k}(-1)^{\epsilon_{m}%
(\epsilon_{k}+1)}. \label{tcov}%
\end{equation}
where $\overset{M}{\nabla}_{i}$ maps a tensor field
$T_{\;\;\;\;\;\;\;\;\ j_{1}\ldots j_{m}}^{i_{1}\ldots i_{n}}(x)$ of type
$(n,m)$ into a tensor field of type $(n,m+1)$ according to%
\begin{align}
T_{\;\;\;\;\;\;\;\;\ j_{1}...j_{m}}^{i_{1}\ldots i_{n}}\overset{M}{\nabla
}_{k}  &  =T_{\;\;\;\;\;\;\;\;\ j_{1}...j_{m},k}^{i_{1}\ldots i_{n}}%
+\underset{r=1}{\overset{n}{\sum}}\,T_{\;\;\;\;\;\;\;\;\;\;\;\;\;\;j_{1}%
...j_{m}}^{i_{1}...l...j_{n}}\,\overset{M}{\Gamma}\,_{\;\;\;lk}^{i_{r}%
}(-1)^{(\epsilon_{i_{r}}+\epsilon_{l})\left(  \epsilon_{l}+\overset
{n}{\underset{p=r+1}{\sum}}\epsilon_{i_{p}}+\underset{p=1}{\overset{m}{\sum}%
}\epsilon_{j_{p}}\right)  }\nonumber\\
&  -\underset{s=1}{\overset{m}{\sum}\,}T_{\;\;\;\;\;\;\;\;\ j_{1}...l...j_{m}%
}^{i_{1}\ldots i_{n}}\,\overset{M}{\Gamma}\,_{\;\,j_{s}k}^{l}(-1)^{(\epsilon
_{j_{s}}+\epsilon_{l})\underset{p=s+1}{\overset{m}{\sum}}\epsilon_{j_{p}}}.
\label{m_nabla}%
\end{align}
In (\ref{tcov}), (\ref{m_nabla}), the functions $\overset{M}{\Gamma
}\,_{\;\;\;ij}^{k}(x)$ are generalized Christoffel symbols (connection
coefficients), having the transformation law%
\[
\overset{M}{\bar{\Gamma}}\,_{\;\;\;ij}^{k}=(-1)^{\epsilon_{j}(\epsilon
_{m}+\epsilon_{i})}\frac{\partial_{r}\bar{x}^{k}}{\partial x^{l}}\overset
{M}{\Gamma}\,_{\;\;mn}^{l}\frac{\partial_{r}x^{n}}{\partial\bar{x}^{j}%
}\frac{\partial_{r}x^{m}}{\partial\bar{x}^{i}}+\frac{\partial_{r}\bar{x}^{k}%
}{\partial x^{m}}\frac{\partial_{r}^{2}x^{m}}{\partial\bar{x}^{i}\partial
\bar{x}^{j}}\,.
\]
In case a local Cartesian system on $M$ does exist, the connection
coefficients $\overset{M}{\Gamma}\,_{\;\;\;ij}^{k}(x)$ possess the property of
(generalized) symmetry:%
\[
\overset{M}{\Gamma}\,_{\;\;ij}^{k}=(-1)^{\epsilon_{i}\epsilon_{j}}\overset
{M}{\Gamma}\,_{\;\;ji}^{k}\,.
\]
With this in mind, the consideration will be restricted to symmetric connections.

Since $x^{i}$ and $\theta_{a}^{i}$ are independent coordinates, expressions
(\ref{tcov}), (\ref{m_nabla}) imply that the vectors $\theta_{a}^{i}$ are
covariantly constant:%
\begin{equation}
\theta_{a}^{i}\overset{\mathcal{M}}{\nabla}_{j}=0. \label{thecov0}%
\end{equation}
From (\ref{tcov}), (\ref{thecov0}), it follows that a (generalized) commutator
$[\overset{\mathcal{M}}{\nabla}_{i},\overset{\mathcal{M}}{\nabla}%
_{j}]=\overset{\mathcal{M}}{\nabla}_{i}\overset{\mathcal{M}}{\nabla}%
_{j}-(-1)^{\epsilon_{i}\epsilon_{j}}\overset{\mathcal{M}}{\nabla}_{j}%
\overset{\mathcal{M}}{\nabla}_{i}$, acting on a scalar field $T\left(
x,\theta\right)  $, can be written as%
\begin{equation}
T[\overset{\mathcal{M}}{\nabla}_{i},\overset{\mathcal{M}}{\nabla}%
_{j}]=(-1)^{\epsilon_{m}(\epsilon_{n}+1)}\frac{\partial_{r}T}{\partial
\theta_{a}^{n}}\theta_{a}^{m}\overset{M}{R}\,_{\;\;mij}^{n}\,, \label{commut0}%
\end{equation}
where $\overset{M}{R}\,_{\;\;mjk}^{i}(x)$ is a curvature tensor on the
supermanifold $M$, defined by the action of a commutator of covariant
derivatives $\overset{M}{\nabla}_{i}$ on a vector field $T^{i}\left(
x\right)  $ according to%
\[
T^{i}[\overset{M}{\nabla}_{j},\overset{M}{\nabla}_{k}]=-(-1)^{\epsilon
_{m}(\epsilon_{i}+1)}T^{m}\overset{M}{R}\,_{\;\;mjk}^{i}\,.
\]
The curvature tensor has the explicit form%
\begin{equation}
\overset{M}{R}\,_{\;\;mjk}^{i}=-\overset{M}{\Gamma}\,_{\;\;mj,k}^{i}%
+\overset{M}{\Gamma}\,_{\;\;mk,j}^{i}(-1)^{\epsilon_{j}\epsilon_{k}}%
+\overset{M}{\Gamma}\,_{\;\;jl}^{i}\overset{M}{\Gamma}\,_{\;\;mk}%
^{l}(-1)^{\epsilon_{j}\epsilon_{m}}-\overset{M}{\Gamma}\,_{\;\;kl}^{i}%
\overset{M}{\Gamma}\,_{\;\;mj}^{l}(-1)^{\epsilon_{k}(\epsilon_{m}+\epsilon
_{j})}, \label{R}%
\end{equation}
and obeys a property of (generalized) antisymmetry and a (super) Jacobi
identity:%
\[
\overset{M}{R}\,_{\;\;mjk}^{i}=-(-1)^{\epsilon_{j}\epsilon_{k}}\overset{M}%
{R}\,_{\;\;mkj}^{i}\,,\;\;\;(-1)^{\epsilon_{j}\epsilon_{l}}\overset{M}%
{R}\,_{\;\;jkl}^{i}+\mathrm{cycle\,}(j,k,l)\equiv0.
\]

In what follows, we call $M$ and $\mathcal{M}$ the base and triplectic
supermanifolds, respectively, and refer to $\overset{\mathcal{M}}{\nabla}_{i}$
as the \emph{triplectic covariant derivative}.

\subsection{Superfield Description}

The complete set of variables arising in various quantization schemes
\cite{BLT,3pl,mod3pl} based on extended BRST\ symmetry can be presented as
$(\phi^{A},\bar{\phi}_{A};\pi_{a}^{A},\phi_{Aa}^{\ast};\lambda^{A}%
,J_{A})=(x^{i},\theta_{a}^{i},y^{i})$, $i=1,2,\ldots$, $N=2n$, $\epsilon
(x^{i})=\epsilon(y^{i})=\epsilon_{i},\;\epsilon(\theta_{a}^{i})=\epsilon
_{i}+1$. This set consists of the field-antifield variables $(\phi^{A}%
,\bar{\phi}_{A},\phi_{Aa}^{\ast})$, Lagrange multipliers $(\pi_{a}^{A}%
,\lambda^{A})$, and sources $J_{A}$ to the fields. In the superfield
formulation \cite{L} of extended BRST symmetry, the variables $(x^{i}%
,\theta_{a}^{i},y^{i})$ are regarded as components of superfields $z^{i}%
(\eta)$ in a superspace with Grassmann coordinates $\eta_{a}$,%
\[
z^{i}{(\eta)=x}^{i}+\eta^{a}\theta_{a}^{i}+\eta^{2}y^{i},\;\;\eta^{2}%
\equiv\frac{1}{2}\eta_{a}\eta^{a}\,,
\]
where raising the $Sp(2)$-indices is performed with the help of the
antisymmetric second-rank tensor $\varepsilon^{ab}$: $\eta^{a}=\varepsilon
^{ab}\eta_{b}$, $\varepsilon^{ac}\varepsilon_{cb}=\delta_{b}^{a}$.

Let us identify the components $(x^{i},\theta_{a}^{i},y^{i})$ with local
coordinates on a supermanifold $\mathcal{N}$, $\mathrm{dim\,}\,\mathcal{N}%
=4N$, where the submanifold with coordinates $(x^{i},\theta_{a}^{i})$ is
chosen as a triplectic supermanifold. At the same time, we define the
transformations of the additional coordinates $y^{i}$, that accompany the
transformations $\left(  x,\theta\right)  \rightarrow(\bar{x},\bar{\theta})$,
to be trivial:%
\[
\bar{x}^{i}=\bar{x}^{i}(x),\;\;\;\bar{\theta}_{a}^{i}=\theta_{a}%
^{j}\frac{\partial\bar{x}^{i}}{\partial{x}^{j}}\,,\;\;\;\bar{y}^{i}=y^{i}.
\]

By analogy with the triplectic supermanifold $\mathcal{M}$, a tensor field of
type $(n,m)$ and rank $n+m$ on the supermanifold $\mathcal{N}$ is defined as
an object which in any local coordinate system $(x,\theta,y)$ is given by a
set of functions $T_{\;\;\;\;\;\;\;\;\ j_{1}\ldots j_{m}}^{i_{1}\ldots i_{n}%
}(z)$ that transform as a tensor field on the base supermanifold $M$. With
this in mind, one can define on $\mathcal{N}$ a superfield extension
$\mathcal{D}_{i}\left(  \eta\right)  $ of the triplectic covariant derivative
$\overset{\mathcal{M}}{\nabla}_{i}$. Namely, one introduces $\mathcal{D}%
_{i}\left(  \eta\right)  $ as an operation that maps a tensor field of type
$(n,m)$ into a tensor field of type $(n,m+1)$ and reduces in a local Cartesian
system on $M$ to the superfield derivative\footnote{As usual, we assume that
$\delta T(z)=\int d^{2}\eta\frac{\partial_{r}T}{\partial z^{i}(\eta)}\delta
z^{i}(\eta)$ and $\int d^{2}\eta=\int d^{2}\eta\;\eta^{a}=0$,$\;\int d^{2}%
\eta\;\eta^{a}\eta^{b}=\varepsilon^{ab}$.}%
\[
\frac{\overleftarrow{\partial}}{\partial z^{i}{(\eta)}}=\frac{\overleftarrow
{\partial}}{\partial x^{i}}\eta^{2}+\frac{\overleftarrow{\partial}}%
{\partial\theta_{a}^{i}}\eta_{a}\,,
\]
defined with respect to variations $\delta z^{i}(\eta)=\delta{x}^{i}+\eta
^{a}\delta\theta_{a}^{i}$ induced by $(x,\theta,y)\rightarrow(\bar{x}%
,\bar{\theta},\bar{y})$. The derivative $\mathcal{D}_{i}\left(  \eta\right)  $
has the explicit form%
\begin{equation}
\overleftarrow{\mathcal{D}}_{i}(\eta)=\overset{\mathcal{M}}{\overleftarrow
{\nabla}}_{i}\,\eta^{2}+\frac{\overleftarrow{\partial}}{\partial\theta_{a}%
^{i}}\,\eta_{a}\,, \label{tcovs}%
\end{equation}
where each term of the $\eta$-expansion transforms as a covector with respect
to $(x,\theta,y)\rightarrow(\bar{x},\bar{\theta},\bar{y})$.

Using $\mathcal{D}_{i}\left(  \eta\right)  $, one can rewrite the equalities
(\ref{thecov0}), (\ref{commut0}) in the form%
\begin{gather}
\frac{\partial z^{i}}{\partial\eta^{a\prime}}\mathcal{D}_{j}\left(
\eta^{\prime\prime}\right)  =\delta_{j}^{i}\eta_{a}^{\prime\prime
},\label{thecov}\\
T\left[  \mathcal{D}_{i}(\eta^{\prime}),\mathcal{D}_{j}(\eta^{\prime\prime
})\right]  =(-1)^{\epsilon_{m}(\epsilon_{n}+1)}(\eta^{\prime})^{2}\left(
\eta^{\prime\prime}\right)  ^{2}\frac{\partial_{r}\left(  T\mathcal{D}%
_{n}\right)  }{\partial\eta_{a}^{\prime}}\frac{\partial z^{m}}{\partial
\eta^{\prime\prime a}}\overset{M}{R}\,_{\;\;mij}^{n}\,, \label{commut}%
\end{gather}
where $T\left(  z\right)  $ is a scalar field, and $\overset{M}{R}%
\,\,_{\;\;mij}^{n}\left(  x\right)  $ is the curvature tensor (\ref{R}).

\section{Extended Superfield Realization of (Modified) Triplectic Algebra}

In this section, we shall apply the above ingredients in order to construct an
extended realization of the triplectic and modified triplectic algebras
\cite{3pl,mod3pl}. To this end, we recall that the triplectic algebra
\cite{3pl} is formed by two doublets of first- and second-order operators,
$\overleftarrow{V}^{a}$ and $\overleftarrow{\Delta}^{a}$, respectively, having
the Grassmann parity $\epsilon(V^{a})=\epsilon(\Delta^{a})=1$ and obeying the
relations%
\begin{equation}
\Delta^{\{a}\Delta^{b\}}=0,\;\;\;V^{\{a}V^{b\}}=0,\;\;\;V^{a}\Delta^{b}%
+\Delta^{b}V^{a}=0, \label{dd}%
\end{equation}
whereas the modified triplectic algebra \cite{mod3pl} involves an additional
doublet of first-order operators $\overleftarrow{U}^{a}$, $\epsilon(U^{a})=1$,
and has the form%
\begin{gather}
\Delta^{\{a}\Delta^{b\}}=0,\;\;\;V^{\{a}V^{b\}}=0,\;\;\;U^{\{a}U^{b\}}%
=0,\nonumber\\
V^{\{a}\Delta^{b\}}+\Delta^{\{b}V^{a\}}=0,\;\;\;\Delta^{\{a}U^{b\}}%
+U^{\{a}\Delta^{b\}}=0,\;\;\;U^{\{a}V^{b\}}+V^{\{a}U^{b\}}=0. \label{mal}%
\end{gather}
In (\ref{dd}), (\ref{mal}), the curly brackets stand for symmetrization with
respect to the enclosed indices.

Using the second-order operators $\Delta^{a}$, one can define a pair of
bilinear operations $(\;\,,\;)^{a}$,%
\begin{equation}
(F,G)^{a}=(-1)^{\epsilon(G)}(FG)\Delta^{a}-(-1)^{\epsilon(G)}(F\Delta
^{a})G-F(G\Delta^{a}). \label{ab}%
\end{equation}
which form a set of antibrackets, similar to those introduced in the
$Sp(2)$-covariant formalism \cite{BLT}. Thus, the operations $(\;\,,\;)^{a}$,
possess the Grassmann parity $\epsilon((F,G)^{a})=\epsilon(F)+\epsilon(G)+1$
and obey the symmetry property%
\[
(F,G)^{a}=-(-1)^{(\epsilon(G)+1)(\epsilon(F)+1)}(G,F)^{a},
\]
as well as the Leibniz rules%
\begin{gather}
(F,GH)^{a}=(F,G)^{a}H+(F,H)^{a}G(-1)^{\epsilon(G)\epsilon(H)},\nonumber\\
(F,G)^{\{a}D^{b\}}=(F,GD^{\{a})^{b\}}-(FD^{\{a},G)^{b\}}(-1)^{\epsilon
(G)},\;\;D^{a}\equiv\{\Delta^{a},U^{a},V^{a}\} \label{leibniz}%
\end{gather}
and the Jacobi identity%
\begin{equation}
(F,(G,H)^{\{a})^{b\}}(-1)^{(\epsilon(F)+1)(\epsilon(H)+1)}+\mathrm{cycle}%
(F,G,H)\equiv0. \label{jacobi}%
\end{equation}

In general coordinates, the operators (\ref{dd}), (\ref{mal}) and the
antibrackets (\ref{ab}) can be constructed \cite{gl,gglm} in terms of a scalar
$\rho\left(  x\right)  $ and tensor $\omega_{ij}\left(  x\right)  $,
$g_{ij}\left(  x\right)  $ fields defined on the base supermanifold $M$. At
the same time, within the superfield description \cite{gglm} the objects
$\rho$, $\omega_{ij}$, $g_{ij}$ are formally identified with fields
$\mathcal{R}$, $\Omega_{ij}$, $G_{ij}$ on the larger supermanifold
$\mathcal{N}$. In contrast to \cite{gglm}, we shall present a superfield
realization of (\ref{dd})--(\ref{ab}) in terms of extended counterparts of
$\rho$, $\omega_{ij}$, $g_{ij}$, inherently defined on $\mathcal{N}$. The
corresponding quantization procedure then follows the approach \cite{gglm}.

\subsection{Extended Realization}

Let us equip the supermanifold $\mathcal{N}$ with an even scalar field
$\mathcal{R}(z)$, as well as with even tensor fields $G_{ij}(z)$ and
$\Omega_{ij}(z)$, the latter having the inverse\footnote{In the supersymmteric
case, the contraction rules for tensor indices, as well as the definition of a
nondegenerate tensor, can be found in the monograph \cite{DeWitt} and papers
\cite{gl,gl2}.} $\Omega^{ij}(z)$, $\epsilon(G_{ij})=\epsilon(\Omega
_{ij})=\epsilon(\Omega^{ij})=\epsilon_{i}+\epsilon_{j}$,%
\begin{equation}
\Omega_{ik}\Omega^{kj}(-1)^{\epsilon_{i}}=\delta_{i}^{j},\;\;\;\Omega
^{ik}\Omega_{kj}(-1)^{\epsilon_{k}}=\delta_{j}^{i}. \label{nongen}%
\end{equation}
We demand that the fields $G_{ij}(z)$ and $\Omega_{ij}(z)$, $\Omega^{ij}(z)$
obey the following properties of generalized (anti)symmetry:%
\[
G_{ij}=(-1)^{\epsilon_{i}\epsilon_{j}}G_{ji},\;\;\Omega_{ij}=-(-1)^{\epsilon
_{i}\epsilon_{j}}\Omega_{ji}\Leftrightarrow\Omega^{ij}=-(-1)^{\epsilon
_{i}\epsilon_{j}}\Omega^{ji}.
\]
At the same time, we require that $\Omega_{ij}(z)$ and $\Omega^{ij}(z)$ be
covariantly constant with respect to the superfield derivative $\mathcal{D}%
_{i}\left(  \eta\right)  $, namely,%
\begin{equation}
\Omega_{ij}\mathcal{D}_{k}=0\Leftrightarrow\Omega^{ij}\mathcal{D}_{k}=0.
\label{2}%
\end{equation}

Using the fields $\mathcal{R}(z)$, $\Omega^{ij}(z)$ and the derivative
$\mathcal{D}_{i}(\eta)$, one can construct a superfield $Sp(2)$-doublet
$\Delta^{a}$ of odd second-order differential operators, acting as scalars on
the supermanifold $\mathcal{N}$,%
\begin{equation}
\overleftarrow{\Delta}^{a}=\int d^{2}\eta\,{\eta}^{2}\frac{\partial
_{r}\overleftarrow{\mathcal{D}}^{i}}{\partial\eta_{a}}\frac{\partial}%
{\partial{\eta}^{2}}\left(  \overleftarrow{\mathcal{D}}_{i}+\frac{1}%
{2}(\mathcal{R}\overleftarrow{\mathcal{D}}_{i})\right)  , \label{Delt}%
\end{equation}
where $\mathcal{D}^{i}(\eta)$ is a superfield derivative,%
\[
\mathcal{D}^{i}=\mathcal{D}_{j}\Omega^{ji}\Leftrightarrow\mathcal{D}%
_{i}=\mathcal{D}^{j}\Omega_{ji}(-1)^{\epsilon_{j}},
\]
which transforms as a vector on $\mathcal{N}$. In accordance with (\ref{ab}),
the operators (\ref{Delt}) generate a doublet of superfield bracket
operations:%
\begin{equation}
(F,G)^{a}=-\int d^{2}\eta\,\eta^{2}\frac{\partial\left(  F\mathcal{D}%
^{i}\right)  }{\partial{\eta}^{2}}\frac{\partial\left(  G\mathcal{D}%
_{i}\right)  }{\partial\eta_{a}}(-1)^{\epsilon_{i}\epsilon(G)}-(-1)^{(\epsilon
(F)+1)(\epsilon(G)+1)}(F\leftrightarrow G). \label{2op}%
\end{equation}

Using the fields $G_{ij}(z)$ and $\Omega_{ij}(z)$, one can also equip the
supermanifold $\mathcal{N}$ with the superfield objects $S_{0}$ and $S_{ab}$%
\begin{align}
S_{0}  &  =\frac{1}{2}\int d^{2}\eta\,\eta^{2}\frac{\partial_{r}z^{i}%
}{\partial\eta_{a}}G_{ij}\frac{\partial_{r}z^{j}}{\partial\eta^{a}%
}\,,\;\;\;\epsilon(S_{0})=0,\nonumber\\
S_{ab}  &  =-\frac{1}{6}\int d^{2}\eta\,{\eta}^{2}\frac{\partial_{r}z^{i}%
}{\partial\eta^{a}}\Omega_{ij}\frac{\partial_{r}z^{j}}{\partial\eta^{b}%
}\,,\;\;\;\epsilon(S_{ab})=0, \label{S0}%
\end{align}
invariant under local coordinate transformations, $\bar{S}_{0}=S_{0}$,
${\bar{S}}_{ab}=S_{ab}$. Here, $S_{0}$ is an $Sp(2)$-scalar, whereas $S_{ab}$
is an $Sp(2)$-tensor, symmetric with respect to its indices, $S_{ab}=S_{ba}$.

Using $S_{0}$, $S_{ab}$ and the bracket operations $(\,\;,\;)^{a}$, we define
the $Sp(2)$-doublets of first-order odd differential operators $V^{a}$ and
$U^{a}$%
\begin{align}
\overleftarrow{V}_{a}=(\;\cdot,S_{ab})^{b}=  &  -\frac{1}{2}\int d^{2}%
\eta\,\eta^{2}\frac{\partial\overleftarrow{\mathcal{D}}_{i}}{\partial\eta^{2}%
}\frac{\partial_{r}z^{i}}{\partial\eta^{a}}\,,\label{Va}\\
\overleftarrow{U}^{a}=(\;\cdot,S_{0})^{a}=  &  \int d^{2}\eta\,\eta
^{2}\left\{  \frac{\partial\overleftarrow{\mathcal{D}}^{i}}{\partial\eta^{2}%
}\left[  G_{ij}\frac{\partial z^{j}}{\partial\eta_{a}}(-1)^{\epsilon_{i}%
}+\frac{1}{2}\frac{\partial_{r}z^{j}}{\partial\eta_{b}}\frac{\partial
_{r}\left(  G_{jk}\overleftarrow{\mathcal{D}}_{i}\right)  }{\partial\eta_{a}%
}\frac{\partial z^{k}}{\partial\eta^{b}}(-1)^{\epsilon_{i}\epsilon_{k}%
}\right]  \right. \nonumber\\
&  +\left.  \frac{1}{2}\frac{\partial_{r}\overleftarrow{\mathcal{D}}^{i}%
}{\partial\eta_{a}}\frac{\partial_{r}z^{j}}{\partial\eta^{b}}\frac{\partial
\left(  G_{jk}\overleftarrow{\mathcal{D}}_{i}\right)  }{\partial\eta^{2}%
}\frac{\partial_{r}z^{k}}{\partial\eta_{b}}(-1)^{\epsilon_{i}\left(
\epsilon_{k}+1\right)  }\right\}  . \label{Ua}%
\end{align}

The objects in (\ref{Delt})--(\ref{Va}) are formally identical with those
arising in \cite{gglm}. At the same time, the doublet of first-order operators
(\ref{Ua}) is an extension of its counterpart from \cite{gglm}; namely, it
contains an extra term with the expression $\partial_{r}\left(  G_{ij}%
\mathcal{D}_{k}\right)  /\partial\eta_{a}$, related to the additional
variables present in $G_{ij}$.

By straightforward calculations (see Appendix A), taking into account the
expressions (\ref{Delt}), (\ref{Va}) for $\Delta^{a}$, $V^{a}$ and the
properties (\ref{thecov}), (\ref{commut}) of derivatives $\mathcal{D}_{i}%
(\eta)$, one can show that the triplectic algebra (\ref{dd}) is fulfilled if
the scalar field $\mathcal{R}(z)$ is chosen as%
\begin{equation}
\mathcal{R}=-\mathrm{log}\;\mathrm{sdet}\;\left(  \Omega^{ij}\right)  ,
\label{rho}%
\end{equation}
while the tensor fields $\Omega_{ij}(z)$ and $\Omega^{ij}(z)$ obey the Jacobi
identities%
\begin{equation}
\frac{\partial_{r}\Omega_{ij}}{\partial z^{k}}(-1)^{\epsilon_{i}\epsilon_{k}%
}+\mathrm{cycle\,}(i,j,k)=0\Leftrightarrow\Omega^{il}\frac{\partial\Omega
^{jk}}{\partial z^{l}}(-1)^{\epsilon_{i}\epsilon_{k}}+\mathrm{cycle\,}%
(i,j,k)=0, \label{3}%
\end{equation}
and the curvature tensor (\ref{R}) of the base supermanifold $M$ is zero:%
\begin{equation}
\overset{M}{R}\,_{\;\;mjk}^{i}=0. \label{4}%
\end{equation}
Now that the triplectic algebra (\ref{dd}) and the related antibrackets
(\ref{ab}) have been explicitly constructed, the Leibniz rules (\ref{leibniz})
and the Jacobi identity (\ref{jacobi}) are obviously fulfilled.

Due to (\ref{dd}), in order to complete the construction of the modified
triplectic algebra (\ref{mal}), it remains to ensure the fulfillment of the
relations involving the operators $U^{a}$. In this respect, the definition
(\ref{Ua}), namely, $\overleftarrow{U}^{a}=(\;\cdot,S_{0})^{a}$, as well as
the Leibniz rules (\ref{leibniz}) and the Jacobi identity (\ref{jacobi}),
imply that the modified triplectic algebra (\ref{mal}) holds true in case the
function $S_{0}$ is subject to%
\begin{equation}
(S_{0},S_{0})^{a}=0,\;\;S_{0}V^{a}=0,\;\;S_{0}\Delta^{a}=0. \label{S1}%
\end{equation}
We note that a solution of these equations can be found (see Appendix B) in
the class of covariantly constant tensor fields $G_{ij}(z)$ with a subsidiary
condition:%
\begin{equation}
G_{ij}\mathcal{D}_{k}=0,\;\;G_{ij}\frac{\partial_{r}\Omega^{jk}}{\partial
z^{k}}(-1)^{k}=0. \label{subsid}%
\end{equation}

The geometric interpretation of equalities (\ref{3}) will be given in the
following subsection, with allowance for (\ref{2}), (\ref{4}). As far as
conditions (\ref{subsid}) are concerned, we do not need to restrict the
consideration to this special case of solutions to (\ref{S1}); namely, in what
follows we merely assume that equations (\ref{S1}) are fulfilled.

\subsection{Quantization Rules}

Having constructed an explicit realization of the differential operators
$\Delta^{a}$, $V^{a}$, $U^{a}$ and antibrackets $(\,\;,\;)^{a}$, we are in a
position to set up a quantization procedure. This procedure repeats the
BRST--antiBRST superfield covariant scheme in general coordinates \cite{gglm}
and has the same features. Thus, the vacuum functional is defined as%
\begin{equation}
Z=\int dz\,\mathcal{D}_{0}\,\mathrm{exp}{\{(i/\hbar)[W+X+\alpha S_{0}]\}},
\label{Z}%
\end{equation}
with $\alpha$ being an arbitrary constant and the function $S_{0}$ given by
(\ref{S0}). The quantum action $W=W(z)$ and the gauge-fixing functional
$X=X(z)$ obey the quantum master equations%
\begin{align}
\frac{1}{2}(W,W)^{a}+W\mathcal{V}^{a}  &  =i\hbar W\Delta^{a},\nonumber\\
\frac{1}{2}(X,X)^{a}+X\mathcal{U}^{a}  &  =i\hbar X\Delta^{a}. \label{MEX}%
\end{align}
Integration in (\ref{Z}) goes over the components of supervariables,
$dz=dx\,d\theta_{a}\,dy$, and the integration measure $\mathcal{D}_{0}$ reads%
\begin{equation}
\mathcal{D}_{0}=\left[  \mathrm{sdet}\left(  \Omega^{ij}\right)  \right]
^{-3/2}. \label{d0}%
\end{equation}
In (\ref{MEX}), we use the operators%
\[
\mathcal{V}^{a}=\frac{1}{2}\left(  \alpha U^{a}+\beta V^{a}+\gamma
U^{a}\right)  ,\;\;\mathcal{U}^{a}=\frac{1}{2}\left(  \alpha U^{a}-\beta
V^{a}-\gamma U^{a}\right)  .
\]
with the properties%

\[
\mathcal{V}^{\{a}\mathcal{V}^{b\}}=0,\;\;\;\mathcal{U}^{\{a}\mathcal{U}%
^{b\}}=0,\;\;\;\mathcal{V}^{\{a}\mathcal{U}^{b\}}+\mathcal{U}^{\{a}%
\mathcal{V}^{b\}}=0,
\]
that hold true for arbitrary values of the constant parameters $\alpha$,
$\beta$, $\gamma$, which implies that the operators $\Delta^{a}$,
$\mathcal{V}^{a}$, $\mathcal{U}^{a}$ also realize the modified triplectic algebra.

The integrand of (\ref{Z}) is invariant under extended BRST transformations,
with the generators%
\[
\delta^{a}=(\;\cdot,W-X)^{a}+\mathcal{V}^{a}-\mathcal{U}^{a},
\]
which implies the independence of the vacuum functional (\ref{Z}) from a
choice of the gauge-fixing function $X$ (for arbitrary $\alpha$, $\beta$,
$\gamma$).

Let us establish the relation between \cite{gglm} and the\textbf{ }present
quantization scheme in more detail. To this end, we introduce the notation%
\[
\Omega_{ij}(z)\equiv\omega_{ij}(x,\theta,y),\;\;G_{ij}(z)\equiv g_{ij}%
(x,\theta,y),\;\;\mathcal{R}(z)\equiv\rho(x,\theta,y),
\]
and examine the special case $\omega_{ij}(x)$, $g_{ij}(x)$, which, in view of
(\ref{rho}), implies $\rho(x)$, so that all the fields $\omega_{ij}$, $g_{ij}%
$, $\rho$ are restricted to the base supermanifold $M$, as in the case of
\cite{gglm}.

From (\ref{2}) and (\ref{3}), it follows that $\omega_{ij}(x)$ and its inverse
$\omega^{ij}(x)$ are subject to%
\begin{gather}
\omega_{ij}\overset{M}{\nabla}_{k}=0\Leftrightarrow\omega^{ij}\overset
{M}{\nabla}_{k}=0,\label{onecomp1}\\
\omega_{ij,k}(-1)^{\epsilon_{i}\epsilon_{k}}+\mathrm{cycle\,}(i,j,k)\equiv
0\Leftrightarrow\omega^{il}\partial_{l}\omega^{jk}(-1)^{\epsilon_{i}%
\epsilon_{k}}+\mathrm{cycle\,}(i,j,k)\equiv0, \label{onecomp2}%
\end{gather}
and, therefore, $\omega_{ij}(x)$ and $\omega^{ij}(x)$ are identified with the
antisymmetric fields of \cite{gglm}. Geometrically, (\ref{onecomp2}) implies
that $\omega_{ij}(x)$ and $\omega^{ij}(x)$ provide the base supermanifold $M$
with an even symplectic structure and with a Poisson bracket, respectively
\cite{gl}. At the same time, (\ref{onecomp1}) ensures the covariant constancy
of the even differential two-form $\omega=\omega_{ij}dx^{j}\wedge dx^{i}$, so
that $M$ is interpreted as an even Fedosov supermanifold \cite{gl,gl2}, i.e.,
an extension of Fedosov manifolds \cite{F,fm} to a supersymmetric case. We
also observe that, due to the subsidiary condition (\ref{4}), the Fedosov
supermanifold $M$ is flat, as in the case of \cite{gglm}.

For the fields $\omega_{ij}(x)$, $g_{ij}(x)$, equalities (\ref{Delt}%
)--(\ref{rho}) and (\ref{d0}) imply that the functions $S_{0}$, $\mathcal{D}%
_{0}$, the operators $\Delta^{a}$, $V^{a}$, $U^{a}$ and the antibrackets
$(\;\,,\;)^{a}$ are reduced to the corresponding ingredients of \cite{gglm}.
In this respect, we note that $U^{a}$ are reduced to their counterparts of
\cite{gglm} due to the equality%
\[
\int d^{2}\eta\,\eta^{2}f(\eta)\frac{\partial\left(  G_{ij}\mathcal{D}%
_{k}\right)  }{\partial\eta_{a}}=0
\]
which holds for $g_{ij}(x)$, with an arbitrary function $f(\eta)$.
Consequently, in the case $\omega_{ij}(x)$, $g_{ij}(x)$, equations (\ref{MEX})
are identical with the master equations of \cite{gglm}, so that expression
(\ref{Z}) reduces to the vacuum functional of \cite{gglm}.

\section{Conclusion}

We have presented a natural extension of the recently proposed BRST--antiBRST
superfield covariant scheme in general coordinates \cite{gglm}. Thus, the
coordinate dependence of the basic ingredients of \cite{gglm}, being the
scalar $\rho$ and tensor $\omega_{ij}$, $g_{ij}$ fields, has been extended to
the complete set of supervariables used in that formalism. In terms of the
extended fields, $\mathcal{R}$, $\Omega_{ij}$, $G_{ij}$, we have explicitly
realized the differential operators $\Delta^{a}$, $V^{a}$, $U^{a}$, subject to
the modified triplectic algebra, and constructed the related antibracket
operations $(\,\;,\;)^{a}$. The corresponding quantization procedure follows
\cite{gglm} and has the same general features. Thus, the formalism contains
free parameters ($\alpha$, $\beta$, $\gamma$), whose arbitrary choice yields a
gauge-independent vacuum functional and, consequently, a gauge-independent
$S$-matrix \cite{t}. In the limit when the extended fields $\mathcal{R}$,
$\Omega_{ij}$, $G_{ij}$ are reduced to the original ingredients of
\cite{gglm}, namely, $\rho$, $\omega_{ij}$, $g_{ij}$, the present quantization
scheme becomes identical with the approach \cite{gglm}, and, therefore,
reproduces (for a specific choice of the parameters $\alpha$, $\beta$,
$\gamma$ in Darboux coordinates \cite{gl,gglm}) the previously known schemes
\cite{BLT,3pl,mod3pl} of covariant BRST--antiBRST quantization. It appears
interesting to combine the considerations of the present work with the ideas
of the recent paper \cite{gln3}, which suggests enlarging the structure of a
Fedosov supermanifold to the case of superfield variables. Such an
opportunity, however, is impeded by the problem of a consistent superfield
extension of connection coefficients \cite{gln3}.

\textbf{Acknowledgments:}~D.M.G. thanks the foundations FAPESP and CNPq for
permanent support. P.Yu.M. and J.L.T. are grateful to FAPESP.

\appendix                           

\section{Triplectic Algebra}

\setcounter{equation}{0}  \renewcommand{\theequation}{A.\arabic{equation}} 

Let us examine the algebra of the differential operators $\Delta^{a}$ and
$V^{a}$, as acting on scalars defined on the supermanifold $\mathcal{N}$.
Using the expressions (\ref{tcov}), (\ref{m_nabla}), (\ref{tcovs}), the
properties (\ref{thecov}), (\ref{commut}), and the definitions (\ref{Delt}),
(\ref{Va}), we obtain%
\begin{align}
\Delta^{\{a}\Delta^{b\}}=  &  -\int d^{2}\eta^{\prime}d^{2}\eta^{\prime\prime
}\left\{  \mathcal{A}_{ij}\left(  \eta^{\prime},\eta^{\prime\prime}\right)
\frac{\partial_{r}\mathcal{D}^{j}}{\partial\eta_{a}^{\prime\prime}%
}\frac{\partial\mathcal{D}^{i}}{\partial\eta_{b}^{\prime}}\right. \nonumber\\
&  +\frac{1}{2}\frac{\partial_{r}\mathcal{D}^{i}}{\partial\eta^{\prime c}%
}\varepsilon^{c\{a}\left[  \mathcal{B}_{i}^{jb\}}\left(  \eta^{\prime}%
,\eta^{\prime\prime}\right)  \frac{\partial\mathcal{D}_{j}}{\partial
\eta^{\prime\prime2}}+\frac{\partial\mathcal{D}_{j}}{\partial\eta
^{\prime\prime2}}\mathcal{B}_{i}^{jb\}}\left(  \eta^{\prime},\eta
^{\prime\prime}\right)  \right. \nonumber\\
&  -\frac{1}{2}\varepsilon^{b\}d}\frac{\partial\mathcal{D}^{j}}{\partial
\eta^{\prime\prime d}}\left[  \mathcal{B}_{ij}\left(  \eta^{\prime}%
,\eta^{\prime\prime}\right)  -\left(  -1\right)  ^{\varepsilon_{i}%
\varepsilon_{j}}\mathcal{B}_{ji}\left(  \eta^{\prime},\eta^{\prime\prime
}\right)  \right] \nonumber\\
&  +\left.  \left.  \frac{1}{2}\mathcal{B}_{i}^{jb\}}\left(  \eta^{\prime
},\eta^{\prime\prime}\right)  \frac{\partial\left(  \mathcal{RD}_{j}\right)
}{\partial\eta^{\prime\prime2}}\right]  \right\}  \eta^{\prime2}\eta
^{\prime\prime2},\nonumber\\
V^{\{a}V^{b\}}=  &  -\frac{1}{4}\int d^{2}\eta^{\prime}d^{2}\eta^{\prime
\prime}\mathcal{A}_{ij}\left(  \eta^{\prime},\eta^{\prime\prime}\right)
\frac{\partial_{r}z^{j}}{\partial\eta_{a}^{\prime\prime}}\frac{\partial z^{i}%
}{\partial\eta_{b}^{\prime}}\eta^{\prime2}\eta^{\prime\prime2},\nonumber\\
2\left(  \Delta^{a}V^{b}+V^{b}\Delta^{a}\right)  =  &  -\int d^{2}\eta
^{\prime}d^{2}\eta^{\prime\prime}\left\{  \mathcal{A}_{ij}\left(  \eta
^{\prime},\eta^{\prime\prime}\right)  \left(  \frac{1}{2}\varepsilon
^{ab}\Omega^{ji}+\frac{\partial_{r}\mathcal{D}^{i}}{\partial\eta_{a}^{\prime}%
}\frac{\partial z^{j}}{\partial\eta_{b}^{\prime\prime}}\right)  \right.
\nonumber\\
&  +\varepsilon^{ab}\frac{\partial\mathcal{D}_{i}}{\partial\eta^{\prime
\prime2}}\frac{\partial}{\partial\eta^{\prime2}}\left[  \frac{\partial
_{r}\Omega^{ij}}{\partial z^{j}(\eta^{\prime})}+\frac{1}{2}\Omega^{ij}\left(
\mathcal{RD}_{j}(\eta^{\prime})\right)  \right]  \left(  -1\right)
^{\varepsilon_{j}}\nonumber\\
&  -\frac{\partial D^{i}}{\partial\eta_{a}^{\prime}}\left(  \frac{1}%
{2}\mathcal{B}_{ij}\left(  \eta^{\prime},\eta^{\prime\prime}\right)
-\overset{M}{\Gamma}\,_{ij,k}^{k}\left(  -1\right)  ^{\varepsilon_{k}\left(
\varepsilon_{i}+\varepsilon_{j}+1\right)  }\right. \nonumber\\
&  \left.  \left.  +\overset{M}{\Gamma}\,_{jl}^{k}\overset{M}{\Gamma}%
\,_{ik}^{l}\left(  -1\right)  ^{\varepsilon_{j}\left(  \varepsilon
_{i}+\varepsilon_{k}\right)  +\varepsilon_{k}\left(  \varepsilon_{i}+1\right)
}\right)  \frac{\partial_{r}z^{j}}{\partial\eta_{b}^{\prime\prime}}\right\}
\eta^{\prime2}\eta^{\prime\prime2}, \label{B3}%
\end{align}
with the following notation:%
\begin{equation}
\mathcal{A}_{ij}\left(  \eta^{\prime},\eta^{\prime\prime}\right)
\equiv\frac{\partial}{\partial\eta^{\prime2}}\frac{\partial}{\partial
\eta^{\prime\prime2}}\left[  \mathcal{D}_{i}(\eta^{\prime}),\mathcal{D}%
_{j}(\eta^{\prime\prime})\right]  ,\;\mathcal{B}_{j}^{ia}\left(  \eta^{\prime
},\eta^{\prime\prime}\right)  \equiv\frac{\partial\left(  \mathcal{RD}%
_{j}\right)  }{\partial\eta^{\prime2}}\frac{\partial_{r}\mathcal{D}^{i}%
}{\partial\eta_{a}^{\prime\prime}},\;\mathcal{B}_{ij}\left(  \eta^{\prime
},\eta^{\prime\prime}\right)  \equiv\frac{\partial\left(  \mathcal{RD}%
_{i}\right)  }{\partial\eta^{\prime2}}\frac{\partial\mathcal{D}_{j}}%
{\partial\eta^{\prime\prime2}}. \label{notn}%
\end{equation}
We now subject the function $\mathcal{R}\left(  z\right)  $ to the equations%
\begin{equation}
\left(  \frac{\partial_{r}\Omega^{ij}}{\partial z^{j}(\eta)}+\frac{1}{2}%
\Omega^{ij}\frac{\partial_{r}\mathcal{R}}{\partial z^{j}(\eta)}\right)
(-1)^{\epsilon_{j}}=0, \label{new1}%
\end{equation}
which, in view of (\ref{nongen}), are equivalent to%
\[
\frac{\partial_{r}\mathcal{R}}{\partial z^{i}(\eta)}=2\frac{\partial_{r}%
\Omega^{jk}}{\partial z^{k}(\eta)}\Omega_{ji}(-1)^{\epsilon_{j}+\epsilon_{k}%
}.
\]
To solve these equations, we use the consequence of the Jacobi identities
(\ref{3})%
\[
\Omega_{kj}\frac{\partial_{r}\Omega^{jk}}{\partial z^{i}(\eta)}%
+2\frac{\partial_{r}\Omega^{jk}}{\partial z^{k}(\eta)}\Omega_{ji}%
(-1)^{\epsilon_{j}+\epsilon_{k}}=0,
\]
and obtain the equality%
\begin{equation}
\frac{\partial_{r}\mathcal{R}}{\partial z^{i}(\eta)}=-\Omega_{kj}%
\frac{\partial_{r}\Omega^{jk}}{\partial z^{i}(\eta)}\,. \label{B8}%
\end{equation}
Thus, the function $\mathcal{R}\left(  z\right)  $ can be chosen as%
\[
\mathcal{R}=-\mathrm{log}\;\mathrm{sdet}\;\left(  \Omega^{ij}\right)  ,
\]
since its variation has the form%
\begin{align}
\delta\mathcal{R}  &  =-\mathrm{log}\;\mathrm{sdet}\;\left(  \Omega
^{ij}+\delta\Omega^{ij}\right)  +\mathrm{log}\;\mathrm{sdet}\;\left(
\Omega^{ij}\right)  =-\mathrm{log}\;\mathrm{sdet}\;\left(  \delta_{j}%
^{i}+(-1)^{\epsilon_{i}}\Omega_{ik}\delta\Omega^{kj}\right) \nonumber\\
&  =-\mathrm{str}\left[  (-1)^{\epsilon_{i}}\Omega_{ik}\delta\Omega
^{kj}\right]  =-\Omega_{ij}\delta\Omega^{ji}. \label{new0}%
\end{align}
From (\ref{tcov}), (\ref{tcovs}), (\ref{2}) and (\ref{new0}), it follows that%
\begin{equation}
\mathcal{RD}_{i}\left(  \eta\right)  -\frac{\partial_{r}\mathcal{R}}{\partial
z^{i}\left(  \eta\right)  }=\Omega_{jk}\frac{\partial_{r}\left(  \Omega
^{kj}\mathcal{D}_{l}\right)  }{\partial\eta_{a}}\frac{\partial z^{m}}%
{\partial\eta^{a}}\eta^{2}\overset{M}{\Gamma}\,_{\;\;mi}^{l}(-1)^{\epsilon
_{m}\left(  \epsilon_{l}+1\right)  }=0\,. \label{new2}%
\end{equation}
Then, due to (\ref{m_nabla}), (\ref{tcovs}), (\ref{2}) and (\ref{B8}), one
obtains%
\begin{align}
\frac{1}{2}\frac{\partial\left(  \mathcal{RD}_{i}\right)  }{\partial\eta^{2}%
}=  &  -\frac{1}{2}\Omega_{jk}\frac{\partial}{\partial\eta^{2}}\left\{
\Omega^{kj}\mathcal{D}_{i}-\eta^{2}\left[  \frac{\partial_{r}\left(
\Omega^{kj}\mathcal{D}_{m}\right)  }{\partial\eta_{a}}\frac{\partial z^{n}%
}{\partial\eta^{a}}\overset{M}{\Gamma}\,_{\;\;ni}^{m}(-1)^{\epsilon_{n}\left(
\epsilon_{m}+1\right)  }\right.  \right. \nonumber\\
&  \left.  \left.  -\Omega^{km}\overset{M}{\Gamma}\,_{\;\;mi}^{j}%
(-1)^{\epsilon_{m}\left(  \epsilon_{j}+1\right)  }-\Omega^{mj}\overset
{M}{\Gamma}\,_{\;\;mi}^{k}(-1)^{\epsilon_{m}\left(  \epsilon_{j}+\epsilon
_{k}+1\right)  +\epsilon_{j}\epsilon_{k}}\right]  \right\}  =\overset
{M}{\Gamma}\,_{\;\;ji}^{j}(-1)^{\epsilon_{j}}. \label{B10}%
\end{align}
In view of (\ref{notn}), the properties (\ref{new2}) and (\ref{B10}) imply%
\begin{align}
&  \mathcal{B}_{j}^{ia}\left(  \eta^{\prime},\eta^{\prime\prime}\right)
\eta^{\prime\prime2}=0,\;\;\mathcal{B}_{ij}\left(  \eta^{\prime},\eta
^{\prime\prime}\right)  -\left(  -1\right)  ^{\varepsilon_{i}\varepsilon_{j}%
}\mathcal{B}_{ji}\left(  \eta^{\prime},\eta^{\prime\prime}\right)
=0,\nonumber\\
&  \mathcal{B}_{ij}\left(  \eta^{\prime},\eta^{\prime\prime}\right)  =2\left(
\overset{M}{\Gamma}\,_{ki,j}^{k}-\overset{M}{\Gamma}\,_{kl}^{k}\overset
{M}{\Gamma}\,_{ij}^{l}\right)  \left(  -1\right)  ^{\epsilon_{k}}. \label{B11}%
\end{align}
Now, with the help of (\ref{R}), (\ref{new1}) and (\ref{new2})--(\ref{B11}),
the expressions for $\Delta^{\{a}\Delta^{b\}}$ and $\Delta^{a}V^{b}%
+V^{b}\Delta^{a}$ in (\ref{B3}) can be written as follows:%
\begin{align}
\Delta^{\{a}\Delta^{b\}}=  &  -\int d^{2}\eta^{\prime}d^{2}\eta^{\prime\prime
}\mathcal{A}_{ij}\left(  \eta^{\prime},\eta^{\prime\prime}\right)
\frac{\partial_{r}\mathcal{D}^{j}}{\partial\eta_{a}^{\prime\prime}%
}\frac{\partial\mathcal{D}^{i}}{\partial\eta_{b}^{\prime}}\eta^{\prime2}%
\eta^{\prime\prime2},\nonumber\\
2\left(  \Delta^{a}V^{b}+V^{b}\Delta^{a}\right)  =  &  -\int d^{2}\eta
^{\prime}d^{2}\eta^{\prime\prime}\mathcal{A}_{ij}\left(  \eta^{\prime}%
,\eta^{\prime\prime}\right)  \left(  \frac{1}{2}\varepsilon^{ab}\Omega
^{ji}+\frac{\partial_{r}\mathcal{D}^{i}}{\partial\eta_{a}^{\prime}%
}\frac{\partial z^{j}}{\partial\eta_{b}^{\prime\prime}}\right)  \eta^{\prime
2}\eta^{\prime\prime2}\nonumber\\
&  +\int d^{2}\eta^{\prime}d^{2}\eta^{\prime\prime}\frac{\partial
\mathcal{D}^{i}}{\partial\eta_{a}^{\prime}}\overset{M}{R}\,_{\;\;ikj}%
^{k}\frac{\partial_{r}z^{j}}{\partial\eta_{b}^{\prime\prime}}\eta^{\prime
2}\eta^{\prime\prime2}\left(  -1\right)  ^{\epsilon_{k}(\epsilon_{i}+1)}.
\label{dv+vd}%
\end{align}
We remind that the commutator of derivatives $\mathcal{D}_{i}(\eta)$, acting
on scalars defined on the supermanifold $\mathcal{N}$, is given by
(\ref{commut}). Then, due to (\ref{B3}) and (\ref{dv+vd}), we can see that if
the base supermanifold $M$ is chosen to be flat, $\overset{M}{R}%
\,_{\;\;jkl}^{i}=0$, the supermanifold $\mathcal{N}$ admits an explicit
realization of the triplectic algebra (\ref{dd}). As a consequence, the
bracket operations (\ref{2op}) obey the properties (\ref{leibniz}) and
(\ref{jacobi}), so that these operations can be interpreted as extended antibrackets.

\section{Modified Triplectic Algebra}

\setcounter{equation}{0}  \renewcommand{\theequation}{B.\arabic{equation}} 

Let us establish conditions that ensure the fulfillment of equations
(\ref{S1}). To this end, we recall that the function $S_{0}$ has the form
(\ref{S0}), whereas the operators $\Delta^{a}$, $V^{a}$ and antibrackets
$(\;\,,\;)^{a}$ are given by (\ref{Delt}), (\ref{2op}), (\ref{Va}). Using the
properties (\ref{thecov}), (\ref{commut}) of derivatives $\mathcal{D}_{i}%
(\eta)$, we obtain%
\begin{align}
(S_{0},S_{0})^{a}=  &  \int d^{2}\eta^{\prime}d^{2}\eta^{\prime\prime
}\frac{\partial_{r}z^{i}}{\partial\eta^{\prime b}}\left(  G_{ij}%
\mathcal{D}^{k}\left(  \eta^{\prime\prime}\right)  \right)  \frac{\partial
_{r}z^{j}}{\partial\eta_{b}^{\prime}}\frac{\partial\left(  S_{0}%
\mathcal{D}_{k}\right)  }{\partial\eta_{a}^{\prime\prime}}\eta^{\prime2}%
\eta^{\prime\prime2}\left(  -1\right)  ^{\varepsilon_{k}\left(  \varepsilon
_{j}+1\right)  },\nonumber\\
S_{0}V^{a}=  &  \frac{1}{4}\int d^{2}\eta^{\prime}d^{2}\eta^{\prime\prime
}\frac{\partial_{r}z^{i}}{\partial\eta^{\prime b}}\frac{\partial\left(
G_{ij}\mathcal{D}_{k}\right)  }{\partial\eta^{\prime\prime2}}\frac{\partial
_{r}z^{j}}{\partial\eta_{b}^{\prime}}\frac{\partial z^{k}}{\partial\eta
_{a}^{\prime\prime}}\eta^{\prime2}\eta^{\prime\prime2}\left(  -1\right)
^{\varepsilon_{j}\varepsilon_{k}},\nonumber\\
S_{0}\Delta^{a}=  &  \int d^{2}\eta^{\prime}d^{2}\eta^{\prime\prime}\left\{
\frac{\partial_{r}z^{i}}{\partial\eta_{a}^{\prime}}\frac{\partial}%
{\partial\eta^{\prime\prime2}}\left[  \left(  G_{ij}\mathcal{D}^{j}\right)
+\frac{1}{2}G_{ij}\left(  \mathcal{RD}^{j}\right)  \right]  \right.
\nonumber\\
&  +\left.  \frac{1}{2}\frac{\partial_{r}z^{i}}{\partial\eta_{b}^{\prime}%
}\frac{\partial_{r}\left(  G_{ij}\mathcal{D}^{k}\right)  }{\partial\eta
_{a}^{\prime\prime}}\frac{\partial z^{j}}{\partial\eta^{\prime b}%
}\frac{\partial}{\partial\eta^{\prime\prime2}}\left[  \mathcal{D}_{k}%
+\frac{1}{2}\left(  \mathcal{RD}_{k}\right)  \right]  (-1)^{\varepsilon
_{k}\left(  \varepsilon_{j}+1\right)  }\right\}  \eta^{\prime2}\eta
^{\prime\prime2}. \label{mod3pl2}%
\end{align}
Next, imposing the condition of covariance $G_{ij}\mathcal{D}_{k}=0$
(equivalent to $G_{ij}\mathcal{D}^{k}=0$) and the subsidiary condition%
\begin{equation}
G_{ij}\left(  \mathcal{RD}^{j}\right)  =0, \label{condr}%
\end{equation}
we can see that the right-hand sides in (\ref{mod3pl2}) turn to zero. The
above equality (\ref{condr}) is formally identical to a subsidiary condition
used in \cite{gglm}, and, due to (\ref{B8}), (\ref{new2}), is equivalent to%
\[
G_{ij}\frac{\partial_{r}\Omega^{jk}}{\partial z^{k}}(-1)^{k}=0.
\]
Hence, the conditions (\ref{subsid}) provide an example of solutions to
equations (\ref{S1}) which ensure the fulfillment of the modified triplectic
algebra (\ref{mal}).


\begin{thebibliography}{9}                                                                                                %

\bibitem {BLT}I.A. Batalin, P.M. Lavrov and I.V. Tyutin, J. Math. Phys.
\textbf{31}, 1487 (1990); \textbf{32}, 532 (1990); \textbf{32}, 2513
(1990);\newline P.M. Lavrov, Mod. Phys. Lett. \textbf{A6}, 2051 (1991);
\textit{Theor. Math. Phys. }\textbf{89}, 1187 (1991).

\bibitem {3pl}I.A. Batalin and R. Marnelius, \textit{Phys. Lett.}
\textbf{B350}, 44 (1995); \textit{Nucl. Phys.} \textbf{B465}, 521
(1996);\newline I.A. Batalin, R. Marnelius and A.M. Semikhatov, \textit{Nucl.
Phys.} \textbf{B446}, 249 (1995).

\bibitem {mod3pl}B. Geyer, D.M. Gitman and P.M. Lavrov, \textit{Mod. Phys.
Lett.} \textbf{A14}, 661 (1999); \textit{Theor. Math. Phys.} \textbf{123}, 813 (2000).

\bibitem {gln1}B. Geyer, P. Lavrov and A. Nersessian, \textit{Phys. Lett.}
\textbf{B512}, 211 (2001).

\bibitem {gln2}B. Geyer, P. Lavrov and A. Nersessian, \textit{Int. J. Mod.
Phys.} \textbf{A17}, 1183 (2002).

\bibitem {gl}B. Geyer and P. Lavrov, \textit{Int. J. Mod. Phys.} \textbf{A19,}
1639 (2004).

\bibitem {F}B.V. Fedosov, \textit{J. Diff. Geom.} \textbf{40}, 213 (1994);
\emph{Deformation Quantization and Index Theory}\textit{ }(Akademie Verlag,
Berlin, 1996).

\bibitem {gl2}B. Geyer and P. Lavrov, \textit{Int. J. Mod. Phys.} \textbf{A19,
}3195 (2004); \textit{Fedosov supermanifolds. II. Normal coordinates;
}hep-th/0406206; \textit{Basic properties of Fedosov supermanifolds, }hep-th/0406236.

\bibitem {L}P.M. Lavrov, \textit{Phys. Lett. }\textbf{B366}, 160 (1996);
\textit{Theor. Math. Phys. }\textbf{107}, 602 (1996);\newline P.M. Lavrov and
P.Yu. Moshin, \textit{Phys. Lett. }\textbf{B508}, 127 (2001); \textit{Theor.
Math. Phys. }\textbf{129}, 1645 (2001).

\bibitem {gglm}B. Geyer, D.M. Gitman, P.M. Lavrov and P.Yu. Moshin, Int. J.
Mod. Phys. \textbf{A19,} 737 (2004).

\bibitem {gln3}B. Geyer, P. Lavrov and A. Nersessian, \textit{A note on the
supersymplectic structure of triplectic formalism,} hep-th/0406201.

\bibitem {DeWitt}B. DeWitt, \textit{Supermanifolds, }2nd ed. (Cambridge
University Press, Cambridge, 1992).

\bibitem {condnot}B.S. DeWitt, \emph{Dynamical Theory of Groups and Fields}
(Gordon and Breach, New York, 1965).

\bibitem {fm}I.\ Gelfand, V.\ Retakh and M.\ Shubin, \textit{Advan. Math.}
\textbf{136}, 104 (1998); dg-ga/9707024.

\bibitem {t}I.V. Tyutin, \textit{Phys. Atom. Nucl.} \textbf{65}, 194 (2002).
\end{thebibliography}
\end{document}